\documentclass[prl,twocolumn,floatfix,a4paper,showpacs]{revtex4}
\usepackage{amsfonts}
\usepackage{makeidx}
\usepackage{amssymb}
\usepackage{amsmath}
\usepackage{graphicx}
\usepackage{graphics}

\begin{document}

\title{Quantum discord as a resource for quantum cryptography}
\author{Stefano Pirandola}
\affiliation{Department of Computer Science, University of York, York YO10 5GH, United
Kingdom}

\begin{abstract}
Quantum discord is the minimal bipartite resource which is needed for a
secure quantum key distribution, being a cryptographic primitive equivalent
to non-orthogonality. Its role becomes crucial in device-dependent quantum
cryptography, where the presence of preparation and detection noise
(inaccessible to all parties) may be so strong to prevent the distribution
and distillation of entanglement. The necessity of entanglement is
re-affirmed in the stronger scenario of device-independent quantum
cryptography, where all sources of noise are ascribed to the eavesdropper.
\end{abstract}

\pacs{03.65.--w, 03.67.--a, 42.50.--p}
\maketitle

\textit{Introduction}.--~One of the hot topics in the quantum information
theory is the quest for the most appropriate measure and quantification of
quantum correlations. For pure quantum states, this quantification is
provided by quantum entanglement~\cite{Nielsen} which is the physical
resource at the basis of the most powerful protocols of quantum
communication and computation~\cite{Tele,Ekert,Shor}. However, we have
recently understood that the characterization of quantum correlations is
much more subtle in the general case of mixed quantum states~\cite%
{Qdiscord,Qdiscord2}.

There are in fact mixed states which, despite being separable, have
correlations so strong to be irreproducible by any classical probability
distribution. These residual quantum correlations are today quantified by
quantum discord~\cite{VedralRMP}, a new quantity which has been studied in
several contexts with various operational interpretations and applications,
including work extraction~\cite{Demon}, quantum state merging~\cite%
{Merging1,Merging2}, remote state preparation~\cite{Remote}, quantum
metrology~\cite{Qmetrology}, discrimination of unitaries~\cite{Gu} and
quantum channel discrimination~\cite{Qillumination}.

In this paper, we identify the basic role of quantum discord in one of the
most practical tasks of quantum information, i.e., quantum key distribution
(QKD)~\cite{Gisin}. The claim that quantum discord must be non-zero to
implement QKD is intuitive. In fact, quantum discord and its geometric
formulation are connected with the concept of non-orthogonality, which is
the essential ingredient for quantum cryptography. That said, it is still
very important to characterize the general framework where discord remains
the only available resource for QKD. Necessarily, this must be a scenario
where key distribution is possible despite entanglement being absent.

Here we show that this general scenario corresponds to device-dependent (or
trusted-device) QKD, which encompasses all realistic protocols where the
noise affecting the devices and apparata of the honest parties is assumed to
be trusted, i.e., not coming from an eavesdropper but from the action of a
genuine environment. This can be preparation noise (e.g., due to
imperfections in the optical switches/modulators or coming from the natural
thermal background at lower frequencies~\cite%
{Filip,Usenko,Weedbrook2010,Weedbrook2012,Weedbrook2014}) as well as
measurement noise and inefficiencies affecting the detectors (which could be
genuine or even added by the honest parties~\cite{Renner2005,Pirandola2009}%
). Such trusted noise may be high enough to prevent the distribution and
distillation of entanglement, but still a secure key\ can be extracted due
to the presence of non-zero discord.

By contrast, if the extra noise in the apparata is not trusted but
considered to be the effect of side-channel attacks~\cite{SideCH}, then we
have to enforce device-independent QKD~\cite{Ekert}. In this more demanding
scenario, quantum discord is still necessary for security but more simply
becomes an upper bound to the coherent information. This means that secure
key distribution becomes just a consequence of entanglement distillation.

\textit{Quantum discord.}--~Discord comes from different quantum extensions
of the classical mutual information. The first is quantum mutual
information, measuring the total correlations between two systems, $A$ and $%
B $, and defined as $I(A,B):=S(A)-S(A|B)$, where $S(A)$ is the entropy of
system $A$, and $S(A|B):=S(AB)-S(B)$ its conditional entropy. The second
extension is $C(A|B):=S(A)-S_{\min}(A|B)$, where $S_{\min}(A|B)$ is the
entropy of system $A$ minimized over an arbitrary measurement on $B$. This
local measurement is generally described by a positive operator valued
measure (POVM) $\{M_{y}\}$, defining a random outcome variable $%
Y=\{y,p_{y}\} $ and collapsing system $A$ into conditional states $%
\rho_{A|y}~$\cite{Nota2}. Thus, we have%
\begin{equation*}
S_{\min}(A|B):=\inf_{\{M_{y}\}}S(A|Y),~S(A|Y)=\sum_{y}p_{y}S(\rho_{A|y}),
\end{equation*}
where the minimization can be restricted to rank-1 POVMs~\cite{VedralRMP}.

The quantity $C(A|B)$ quantifies the classical correlations between the two
systems, corresponding to the maximal common randomness achievable by local
measurements and one-way classical communication (CC)~\cite{Winter}. Thus,
quantum discord is defined as the difference between total and classical
correlations~\cite{Qdiscord,Qdiscord2,VedralRMP}%
\begin{equation*}
D(A|B):=I(A,B)-C(A|B)=S_{\min}(A|B)-S(A|B)\geq0.
\end{equation*}
An equivalent formula can be written by noticing that $I_{c}(A\rangle
B):=-S(A|B)$ is the coherent information~\cite{Qcap,Qcap2}. Then,
introducing an ancillary system $E$ which purifies $\rho_{AB}$, we can apply
the Koashi-Winter relation~\cite{Koashi} and write $S_{\min}(A|B)=E_{f}(A,E)$%
, where the latter is the entanglement of formation between $A$ and $E$.
Therefore%
\begin{equation}
D(A|B)=I_{c}(A\rangle B)+E_{f}(A,E)\geq\max\{0,I_{c}(A\rangle B)\}.
\label{KWvar}
\end{equation}

It is important to note that $D(A|B)$ is different from $D(B|A)$, where
system $A$ is measured. For instance, in classical-quantum states $%
\rho_{AB}=\sum _{x}p_{x}|x\rangle_{A}\langle x|\otimes\rho_{B}(x)$, where $A$
embeds a classical variable via the orthonormal set $\{|x\rangle\}$ and $B$
is prepared in non-orthogonal states $\{\rho_{B}(x)\}$, we have $D(B|A)=0$
while $D(A|B)>0$. By contrast, for quantum-classical states ($B$ embedding a
classical variable), we have the opposite situation, i.e., $D(A|B)=0$ and $%
D(B|A)>0$.

\begin{figure}[ptbh]
\vspace{-1.6cm}
\par
\begin{center}
\includegraphics[width=0.50\textwidth] {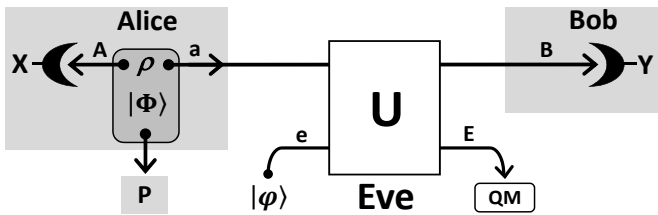}
\end{center}
\par
\vspace{-2.3cm}
\caption{Device-dependent\ protocol with preparation noise. Alice prepares a
generally-mixed input state $\protect\rho_{Aa}$, which is purified into a
state $\Phi_{PAa}$ by adding an extra system $P$ inaccessible to all
parties. System $a$ is sent through an insecure line, so that Alice and Bob
share an output state $\protect\rho_{AB}$. By applying rank-1 POVMs on their
local systems, $A$ and $B$, they derive two correlated random variables, $X$
and $Y$, which are processed into a secret key. In the middle, Eve attacks
the line using a unitary $U$ which couples system $a$ with a pure-state
ancilla $e$. The output ancilla $E$ is then stored in a quantum memory,
which is coherently detected at the end of the protocol.}
\label{picU}
\end{figure}

\textit{Device-dependent QKD protocols.--~}Any QKD\ protocol can be recast
into a measurement-based scheme, where Alice sends Bob part of a bipartite
state, then subject to local detections. Adopting this representation, we
consider a device-dependent protocol where extra noise affects Alice's state
preparation, as in Fig.~\ref{picU} (this is generalized later). In her
private space, Alice prepares two systems, $A$ and $a$, in a generally mixed
state $\rho_{Aa}$. This state is purified into a 3-partite state $\Phi_{PAa}$
with the ancillary system $P$ being inaccessible to Alice, Bob or Eve.

System $a$ is then sent to Bob, who gets the output system $B$. From the
shared state $\rho_{AB}$, Alice and Bob extract two correlated variables:
System $A$ is detected by a rank-1 POVM $\{M_{x}\}$, providing Alice with
variable $X=\{x,p_{x}\}$, while $B$ is detected by another rank-1 POVM $%
\{M_{y}\}$, providing Bob with variable $Y=\{y,p_{y}\}$, whose correlations
with $X$ are quantified by the classical mutual information $I(X,Y)$.

After the previous process has been repeated many times, Alice and Bob
publicly compare a subset of their data. If the error rate is below a
certain threshold, they apply classical procedures of error correction and
privacy amplification with the help of one-way CC, which can be either
forward from Alice to Bob (direct reconciliation), or backward from Bob to
Alice (reverse reconciliation). Thus, they finally extract a secret key at a
rate $K\leq I(X,Y)$, which is denoted by $K(Y|X)$ in direct reconciliation
and $K(X|Y)$ in reverse reconciliation.

To quantify these rates, we need to model Eve's attack. The most general
attack is greatly reduced if Alice and Bob perform random permutations on
their classical data~\cite{Renner,Renner2}. As a result, Eve's attack
collapses into a collective attack, where each travelling system is probed
by an independent ancilla. This means that Eve's interaction can be
represented by a two-system unitary $U_{ae}$ coupling system $a$ with an
ancillary system $e $ prepared in a pure state~\cite{Noteiso}. The output
ancilla $E$\ is then stored in a quantum memory which is coherently measured
at the end of the protocol (see Fig.~\ref{picU}). In this attack, the
maximum information which is stolen on $X$ or $Y$ cannot exceed the Holevo
bound.

\textit{Non-zero discord is necessary}.--~Before analyzing the secret-key
rates, we briefly clarify why discord is a necessary resource for QKD.
Suppose that Alice prepares a quantum-classical state $\rho_{Aa}=%
\sum_{k}p_{k}\rho _{A}(k)\otimes|k\rangle_{a}\langle k|$ with $\{|k\rangle\}$
orthogonal, so that $D(A|a)=0$. Classical system $a$ is perfectly clonable
by Eve. This implies that the three parties will share the state~\cite{Add}%
\begin{equation*}
\rho_{ABE}=\sum_{k}p_{k}\rho_{A}(k)\otimes|k\rangle_{B}\langle k|\otimes
|k\rangle_{E}\langle k|,
\end{equation*}
with Eve fully invisible, since her action is equivalent to an identity
channel for Alice and Bob ($\rho_{AB}=\rho_{Aa}$).

Direct reconciliation fails since $\rho_{ABE}$ is symmetric under $B$-$E$
permutation, which means that Eve decodes Alice's variable with the same
accuracy of Bob. Reverse reconciliation also fails. Bob encodes $Y$ in the
joint state $\rho_{AE|y}=\sum_{k}p_{k|y}\rho_{A}(k)\otimes|k\rangle_{E}%
\langle k|$, where $p_{k|y}:=\langle k|M_{y}|k\rangle$. Then, Eve retrieves $%
K=\{k,p_{k|y}\}$ by a projective POVM, while Alice decodes a variable $X$
with distribution
\begin{equation*}
p_{x|y}=\mathrm{Tr}(M_{x}\rho_{A|y})=\sum_{k}p_{x|k}p_{k|y},~p_{x|k}:=%
\mathrm{Tr}[M_{x}\rho_{A}(k)].
\end{equation*}
This equation defines a Markov chain $Y\rightarrow K\rightarrow X$, so that $%
I(Y,K)\geq I(Y,X)$ by data processing inequality, i.e., Eve gets more
information than Alice~\cite{cohe}.

As expected, system $a$ sent through the channel must be quantum $D(A|a)>0$
in order to have a secure QKD. Indeed, this is equivalent to sending an
ensemble of non-orthogonal states. By contrast, the classicality of the
private system $A$ is still acceptable, i.e., we can have $D(a|A)=0$. In
fact, we may build QKD protocols with preparation noise using
classical-quantum states
\begin{equation}
\rho_{Aa}=\sum_{x}p_{x}|x\rangle_{A}\langle x|\otimes\rho_{a}(x),
\label{cqSTATE}
\end{equation}
whose local detection (on system $A$) prepares any desired ensemble of
non-orthogonal signal states $\{\rho_{a}(x),p_{x}\}$. For instance, the
classical-quantum state of two qubits $\rho_{Aa}=(|0,0\rangle_{Aa}%
\langle0,0|+|1,\varphi\rangle_{Aa}\langle1,\varphi|)/2$, with $\{|0\rangle
,|1\rangle\}$ orthonormal and $\langle0|\varphi\rangle\neq0$, realizes the
B92 protocol~\cite{B92}.

\begin{figure}[ptbh]
\vspace{-2.1cm}
\par
\begin{center}
\includegraphics[width=0.50\textwidth] {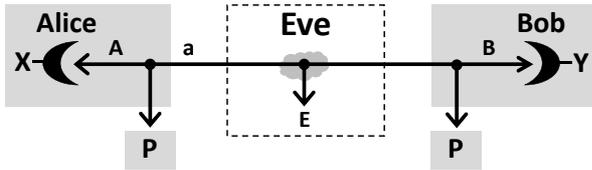}
\end{center}
\par
\vspace{-2.3cm}
\caption{Output state from a device-dependent QKD protocol. Alice and Bob
extract a secret-key by applying rank-1 POVMs on their local systems $A$ and
$B$. Eve steals information from system $E$, while the extra system $P$ is
inaccessible and completes the purification of the global state $\Psi_{ABEP}$%
.}
\label{PicD}
\end{figure}

\textit{Secret-key rates}.--~Once we have clarified that non-zero input
discord $D(A|a)>0$\ is a necessary condition for QKD, we now study the
secret-key rates which can be achieved by device-dependent protocols. Our
next derivation refers to the protocol of Fig.~\ref{picU} and, more
generally, to the scheme of Fig.~\ref{PicD}, where Alice and Bob share an
output state $\rho_{AB}$, where only part of the purification is accessible
to Eve (system $E$), while the inaccessible part $P$ accounts for all
possible forms of extra noise in Alice's and Bob's apparata, including
preparation noise and detection noise (quantum inefficiencies, etc...)

In direct reconciliation, Alice's variable $X$ is the encoding to guess. The
key rate is then given by $K(Y|X)=I(X,Y)-I(E,X)$, where $I(E,X)=S(E)-S(E|X)$
is the Holevo bound quantifying the maximal information that Eve can steal
on Alice's variable~\cite{Notation1}. We can write an achievable upper bound
if we allow Bob to use a quantum memory and a coherent detector. In this
case, $I(X,Y)$ must be replaced by the Holevo quantity $I(B,X)=S(B)-S(B|X)$
and we get the forward Devetak-Winter (DW) rate~\cite{DW1,DW2,DWnote}%
\begin{equation}
K(Y|X)\leq K(B|X):=I(B,X)-I(E,X).   \label{DWdr}
\end{equation}
The optimal forward-rate is defined by optimizing on Alice's individual
detections $K(\blacktriangleright):=\sup_{\{M_{x}\}}K(B|X)$.

We can write similar quantities in reverse reconciliation, where Bob's
variable $Y$ is the encoding to infer. The secret key rate is given by $%
K(X|Y)=I(X,Y)-I(E,Y)$, where $I(E,Y)=S(E)-S(E|Y)$ is Eve's Holevo
information on $Y$. Assuming a coherent detector for Alice, this rate is
bounded by the backward DW rate%
\begin{equation}
K(X|Y)\leq K(A|Y):=I(A,Y)-I(E,Y),   \label{DWrevEQ}
\end{equation}
which gives the optimal backward-rate $K(\blacktriangleleft)$ by maximizing
on Bob's individual detections $\{M_{y}\}$.

Playing with system $P$, we can easily derive upper and lower bounds for the
two optimal rates. Clearly, we get lower bounds $K_{\ast}\leq K$\ if we
assume $P$ to be accessible to Eve, which means to extend $E$ to the joint
system $\mathbf{E}=EP$ in previous Eqs.~(\ref{DWdr}) and~(\ref{DWrevEQ}). By
exploiting the purity of the global state $\Psi_{AB\mathbf{E}}$ and the fact
that the encoding detections are rank-1 POVMs (therefore collapsing pure
states into pure states), we can write the entropic equalities $S(AB)=S(%
\mathbf{E})$, $S(B|X)=S(\mathbf{E}|X)$ and $S(A|Y)=S(\mathbf{E}|Y)$. Then we
easily derive%
\begin{equation*}
K_{\ast}(\blacktriangleright)=I_{c}(A\rangle B),~K_{\ast}(\blacktriangleleft
)=I_{c}(B\rangle A),
\end{equation*}
where the coherent information $I_{c}(A\rangle B)$ and its reverse
counterpart~\cite{RevCOH,RMP} $I_{c}(B\rangle A)$ quantify the maximal
entanglement which is distillable by local operations and one-way CC,
forward and backward, respectively.

It is also clear that we get upperbounds $K^{\ast}\geq K$ by assuming $P$ to
be accessible to the decoding party, Alice or Bob, depending on the
reconciliation. In direct reconciliation, we assume $P$ to be accessible to
Bob, which means extending his system $B$ to $\mathbf{B}=BP$ in Eq.~(\ref%
{DWdr}). Using the equalities $S(A\mathbf{B})=S(E)$ and $S(\mathbf{B}%
|X)=S(E|X)$, we get
\begin{equation*}
K^{\ast}(\blacktriangleright)=I_{c}(A\rangle\mathbf{B})=I_{c}(A\rangle
B)+I(A,P|B),
\end{equation*}
where $I(A,P|B)\geq0$ is the conditional quantum mutual information. Then,
in reverse reconciliation, we assume $P$ to be accessible to Alice, so that $%
A$ becomes $\mathbf{A}=AP$ in Eq.~(\ref{DWrevEQ}). Using $S(\mathbf{A}%
B)=S(E) $ and $S(\mathbf{A}|Y)=S(E|Y)$, we get $K^{\ast}(\blacktriangleleft
)=I_{c}(B\rangle A)+I(B,P|A)$.

Thus, the optimal key rates satisfy the inequalities%
\begin{align}
I_{c}(A\rangle B)~ & \leq~K(\blacktriangleright)~\leq~I_{c}(A\rangle
B)+I(A,P|B),  \label{eq1} \\
I_{c}(B\rangle A)~ & \leq~K(\blacktriangleleft)~\leq~I_{c}(B\rangle
A)+I(B,P|A),   \label{eq2}
\end{align}
where the right hand sides can be bounded using%
\begin{equation*}
I(A,P|B),I(B,P|A)\leq I(AB,P)\leq2\min\{S(P),S(AB)\}.
\end{equation*}
According to Eqs.~(\ref{eq1}) and~(\ref{eq2}), key distribution can occur ($%
K\geq0$) even in the absence of distillable entanglement ($I_{c}=0$). It is
now important to note the following facts:

(i)~Device-dependent QKD is the only scenario where this is possible. In
fact, only in the presence of trusted noise, i.e., $S(P)>0$, we can have $%
K\geq I_{c}$ in the previous equations. Therefore device-dependent QKD is
the only scenario where security may be achieved in the absence of
distillable entanglement and, more strongly, in the complete absence of
entanglement.

(ii)~In device-dependent QKD, we can indeed build protocols which are secure
($K>0$) despite entanglement being completely absent (in any form,
distillable or bound). Clearly, this is possible as long as the minimal
condition $D(A|a)>0$ is satisfied. A secure protocol based on separable
Gaussian states is explictly shown in the supplementary material.

In general, there is an easy way to design device-dependent protocols which
are secure and free of entanglement. Any prepare and measure\ protocol whose
security is based on the transmission of non-orthogonal states $\{\rho
_{a}(x),p_{x}\}$ can be recast into a device-dependent protocol, which is
based on a classical-quantum state $\rho_{Aa}$ as in Eq.~(\ref{cqSTATE}),
whose classical part $A$ is detected while the quantum part $a$ is sent
through the channel. This is as secure as the original one as long as the
purification of the classical-quantum state is inaccessible to Eve. Thus in
such assumption of trusted noise, any prepare and measure protocol has an
equivalent discord-based representation, where non-zero discord guarantees
security in the place of non-orthogonality.

\textit{Side-channels and device-independent QKD}.--~Let us consider the
more demanding scenario where all sources of noise are untrusted. This means
that the extra noise in Alice's and Bob's apparata comes from side-channel
attacks, i.e., system $P$ in Fig.~\ref{PicD} is controlled by Eve. In this
case, the secret-key rates are given by%
\begin{equation}
K(\blacktriangleright)=I_{c}(A\rangle
B),~K(\blacktriangleleft)=I_{c}(B\rangle A),   \label{standard}
\end{equation}
so that QKD is equivalent to entanglement distillation.

It is easy to check that quantum discord upperbounds these key rates.
Applying Eq.~(\ref{KWvar}) to Eq.~(\ref{standard}), we obtain the
cryptographic relations%
\begin{align}
K(\blacktriangleright) & =D(A|B)-E_{f}(A,E)\leq D(A|B),  \label{B1} \\
K(\blacktriangleleft) & =D(B|A)-E_{f}(B,E)\leq D(B|A).   \label{B2}
\end{align}
The optimal forward rate $K(\blacktriangleright)$, where Alice's variable
must be inferred, equals the difference between the output discord $D(A|B)$,
based on Bob's detections, and the entanglement of formation $E_{f}(A,E)$
between Alice and Eve. Situation is reversed for the other rate $%
K(\blacktriangleleft )$. Note that quantum discord not only provides an
upper bound to the key rates, but its aymmetric definition, $D(A|B)$ or $%
D(B|A)$, is closely connected with the reconciliation direction (direct $%
\blacktriangleright$\ or reverse $\blacktriangleleft$).

\textit{Ideal QKD protocols}.--~In practical quantum cryptography, extra
noise is always present, and we distinguish between device-dependent and
device-independent QKD on the basis of Eve's accessibility of the extra
system $P$. In theoretical studies of quantum cryptography, it is however
common to design and assess new protocols by assuming no-extra noise in
Alice's and Bob's apparata (perfect state preparation and perfect
detections).

This is an ideal scenario where system $P$ of Fig.~\ref{PicD} is simply
absent. For such ideal QKD\ protocols, the secret-key rates satisfy again
Eqs.~(\ref{standard}), (\ref{B1}), and~(\ref{B2}), computed on the
corresponding output states. Remarkably, the discord bound can be found to
be tight in reverse reconciliation. In fact, as we show in the supplementary
material, we can have $K(\blacktriangleleft)=D(B|A)$\ in an ideal protocol
of continuous-variable QKD, where Alice transmits part of an
Einstein-Podolsky-Rosen (EPR) state over a pure-loss channel, such as an
optical fiber.

\bigskip

\textit{Conclusion and discussion}.--~Quantum discord can be regarded as a
bipartite formulation of non-orthogonality, therefore capturing the minimal
requisite for QKD. In this paper we have identified the general framework,
device-dependent QKD, where discord remains the ultimate cryptographic
primitive able to guarantee security in the place of quantum entanglement.

We have considered a general form of device-dependent protocol, where Alice
and Bob share a bipartite state which can be purified by two systems: One
system ($E$) is accessible to Eve, while the other ($P$) is inaccessible and
accounts from the presence of trusted noise, e.g., coming from imperfections
in the state preparation and/or the quantum detections. This is a scenario
where the optimal key rate may outperform the coherent information and key
distribution may occur in the complete absence of entanglement (in any form,
distillable or bound) as long as discord is non-zero. As a matter of fact,
any prepare and measure QKD protocol whose security is based on
non-orthogonal quantum states can be recast into an entanglement-free
device-dependent form which is based on a classical-quantum state, with
non-zero discord transmitted through the channel.

This discord-based representation is secure as long as the extra system $P$
is truly inaccessible to Eve, i.e., Alice's and Bob's private spaces cannot
be accessed. Such a condition fails assuming side-channel attacks, where no
noise can be trusted and $P$ becomes part of Eve's systems. In this case,
the secret-key rates are again dominated by the coherent information, which
means that entanglement remains the crucial resource for device-independent
QKD. For both device-independent QKD and ideal QKD\ (where system $P$ is
absent), discord still represents an upper bound to the optimal secret-key
rates achievable in direct or reverse reconciliation, with non trivial cases
where this bound becomes tight.

In conclusion, quantum discord is a necessary resource for secure QKD. This
is particularly evident in device-dependent QKD where entanglement is a
sufficient but not a necessary resource. Entanglement becomes necessary in
device-independent and ideal QKD, where discord still provides an upper
bound to the secret-key rates. Future work may involve the derivation of a
direct mathematical relation between the amount of quantum discord in Alice
and Bob's output state and the optimal secret-key rates which are achievable
in device-dependent QKD.

\textit{Acknowledgments}.--~S.P. was supported by a Leverhulme Trust
research fellowship and EPSRC (grants EP/J00796X/1 and EP/L011298/1).

\setcounter{section}{0} \setcounter{subsection}{0} \renewcommand{%
\bibnumfmt}[1]{[S#1]} \renewcommand{\citenumfont}[1]{S#1}

\begin{center}
{\huge Supplementary Material}
\end{center}

\section{Keys from separable Gaussian states}

Here we provide a simple example of device-dependent QKD\ protocol which is
based on the distribution of a bipartite Gaussian state which is mixed and
separable (not in a tensor-product, therefore having non-zero discord). We
show that the key rates can be positive despite no entanglement being
present. The reader not familiar with the formalism of bosonic systems and
Gaussian states can find these concepts in Ref.~\cite{RMP2}, whose notation
is here adopted ($\hslash=2$ and unit vacuum noise).

Let us consider a continuous variable QKD\ protocol where Alice prepares two
bosonic modes, $A$ and $a$, in a separable Gaussian state $\rho_{Aa}$, with
zero mean and covariance matrix (CM)%
\begin{equation*}
\mathbf{V}_{Aa}=\left(
\begin{array}{cc}
\mu\mathbf{I} & \mathbf{G} \\
\mathbf{G} & \mu\mathbf{I}%
\end{array}
\right) ,
\end{equation*}
where $\mu\geq1$ and $\mathbf{G}$ is a diagonal correlation block which can
be in one of the following forms%
\begin{equation*}
\mathbf{G}=\left(
\begin{array}{cc}
g &  \\
& g%
\end{array}
\right) :=g\mathbf{I,~G}=\left(
\begin{array}{cc}
g &  \\
& -g%
\end{array}
\right) :=g\mathbf{Z}.
\end{equation*}
Here the parameter $g$ must satisfy $|g|\leq\mu-1$, so that $\mathbf{V}_{Aa}$
is both physical and separable~\cite{twomodes}. Apart from the singular case
$g=0$, this symmetric Gaussian state has always non-zero discord, i.e., $%
D(A|a)=D(a|A)>0$~\cite{VeriDISCO}.

Mode $a$ is sent through the channel, where Eve performs a collective
Gaussian attack, whose most general description can be found in Ref~\cite%
{colleGAUSS}. Assuming random permutations (so that quantum de Finetti
applies), this is the most powerful attack against Gaussian protocols~\cite%
{RMP2}. One of the canonical forms of this attack is the so-called
`entangling cloner'\ attack~\cite{RMP2}, where Eve uses a beam splitter with
transmissivity $\tau$ to mix the incoming mode $a$ with one mode $e$ of an
EPR state $\rho_{eE^{\prime}}$ with CM
\begin{equation}
\mathbf{V}_{eE^{\prime}}=\left(
\begin{array}{cc}
\omega\mathbf{I} & \sqrt{\omega^{2}-1}\mathbf{Z} \\
\sqrt{\omega^{2}-1}\mathbf{Z} & \omega\mathbf{I}%
\end{array}
\right) :=\mathbf{V}(\omega),   \label{Vomega}
\end{equation}
where $\omega\geq1$. One output mode $B$ is sent to Bob, while the other
output mode $E$ is stored in a quantum memory together with the retained
mode $E^{\prime}$. Such memory will be coherently detected at the end of the
protocol.

In order to extract two correlated (complex) variables, $X$ and $Y$, Alice
and Bob heterodyne their local modes $A$ and $B$. (Note that other protocols
involving homodyne detection for one of the parties or even two homodynes
may be considered as well.) One can easily check that Alice remotely
prepares thermal states on mode $a$. In fact, by heterodyning mode $A$, the
other mode $a$ is collapsed in a Gaussian state $\rho_{a|X}$ with CM $%
\mathbf{V}_{a|X}=(1+\varepsilon)\mathbf{I}$, where%
\begin{equation*}
\varepsilon:=\mu-1-\frac{g^{2}}{\mu+1}\geq0
\end{equation*}
quantifies the thermalization above the coherent state. This conditional
thermal state is randomly displaced in the phase space according to a
bivariate Gaussian distribution with variance $\mu-1-\varepsilon$ (so that
the average input state on mode $a$ is thermal with the correct CM $\mu%
\mathbf{I}$).

At the output of the channel, Bob's average state is thermal with CM $\nu
_{B}\mathbf{I}$, where $\nu_{B}:=\tau\mu+(1-\tau)\omega$. By propagating the
conditional thermal state $\rho_{a|X}$, we also get Bob's conditional state $%
\rho_{B|X}$, which is randomly displaced and has CM $\nu_{B|X}\mathbf{I}$,
where%
\begin{equation*}
\nu_{B|X}:=\tau(1+\varepsilon)+(1-\tau)\omega=\nu_{B}-\frac{\tau g^{2}}{\mu
+1}.
\end{equation*}
Therefore, we can easily compute Alice and Bob's mutual information, which
is equal to
\begin{equation*}
I(X,Y)=\log_{2}\frac{\nu_{B}+1}{\nu_{B|X}+1}.
\end{equation*}

The next step is the calculation of Eve's Holevo information on Alice's and
Bob's variables. We derive\ the global state of Alice, Bob and Eve, which is
pure Gaussian with zero mean and CM%
\begin{equation*}
\mathbf{V}_{ABEE^{\prime}}=\left(
\begin{array}{cccc}
\mu\mathbf{I} & \sqrt{\tau}\mathbf{G} & -\sqrt{1-\tau}\mathbf{G} & \mathbf{0}
\\
\sqrt{\tau}\mathbf{G} & \nu_{B}\mathbf{I} & \gamma\mathbf{I} & \delta
\mathbf{Z} \\
-\sqrt{1-\tau}\mathbf{G} & \gamma\mathbf{I} & \nu_{E}\mathbf{I} & \kappa%
\mathbf{Z} \\
\mathbf{0} & \delta\mathbf{Z} & \kappa\mathbf{Z} & \omega\mathbf{I}%
\end{array}
\right) ,
\end{equation*}
where $\mathbf{0}$ is the $2\times2$ zero matrix, and%
\begin{align*}
\nu_{E} & :=\tau\omega+(1-\tau)\mu,~\gamma:=\sqrt{\tau(1-\tau)}(\omega -\mu),
\\
\delta & :=\sqrt{1-\tau}\sqrt{\omega^{2}-1},~\kappa:=\sqrt{\tau(\omega
^{2}-1)}.
\end{align*}

From this global CM, we extract Eve's reduced CM $\mathbf{V}_{EE^{\prime}}:=%
\mathbf{V}_{\mathbf{E}}$ describing the two output modes $\mathbf{E}%
=EE^{\prime}$ of the entangling cloner. This reduced CM\ has symplectic
spectrum~\cite{RMP2}%
\begin{equation*}
\nu_{\mathbf{E}}^{\pm}=\frac{\sqrt{\alpha^{2}+4\beta}\pm\alpha}{2},
\end{equation*}
where $\alpha:=(1-\tau)(\mu-\omega)$ and $\beta:=\tau+(1-\tau)\mu\omega$.
The von Neumann entropy of Eve's average state is then given by $S(\mathbf{E}%
)=h(\nu_{\mathbf{E}}^{+})+h(\nu_{\mathbf{E}}^{-})$, where%
\begin{equation*}
h(x):=\frac{x+1}{2}\log_{2}\frac{x+1}{2}-\frac{x-1}{2}\log_{2}\frac{x-1}{2}.
\end{equation*}

By transforming the global CM under heterodyne detection~\cite{RMP2}, we
compute Eve's conditional CMs. First, we derive Eve's CM conditioned to
Bob's detection
\begin{equation*}
\mathbf{V}_{\mathbf{E}|Y}=\mathbf{V}_{\mathbf{E}}-\frac{1}{\nu_{B}+1}\left(
\begin{array}{cc}
\gamma^{2}\mathbf{I} & \gamma\delta\mathbf{Z} \\
\gamma\delta\mathbf{Z} & \delta^{2}\mathbf{I}%
\end{array}
\right) ,
\end{equation*}
which has symplectic spectrum%
\begin{equation*}
\nu_{\mathbf{E}|Y}^{-}=1,~\nu_{\mathbf{E}|Y}^{+}=\frac{\mu+\beta}{1+\mu
\tau+(1-\tau)\omega}.
\end{equation*}
Then, Eve's CM conditioned to Alice's detection is
\begin{equation*}
\mathbf{V}_{\mathbf{E}|X}=\mathbf{V}_{\mathbf{E}}-\frac{(1-\tau)g^{2}}{\mu +1%
}\left(
\begin{array}{cc}
\mathbf{I} & \mathbf{0} \\
\mathbf{0} & \mathbf{0}%
\end{array}
\right) ,
\end{equation*}
and has symplectic spectrum%
\begin{equation*}
\nu_{\mathbf{E}|X}^{\pm}=\frac{\sqrt{\theta^{2}+4(\mu+1)\phi}\pm\theta}{%
2(\mu+1)},
\end{equation*}
where%
\begin{equation*}
\theta:=(1-\tau)g^{2}-(\mu+1)\alpha,~\phi:=(\mu+1)\beta-(1-\tau)\omega
g^{2}.
\end{equation*}

From the previous conditional spectra, we compute Eve's conditional entropies%
\begin{equation*}
S(\mathbf{E}|X)=h(\nu_{\mathbf{E}|X}^{+})+h(\nu_{\mathbf{E}|X}^{-}),~S(%
\mathbf{E}|Y)=h(\nu_{\mathbf{E}|Y}^{+}),
\end{equation*}
and, therefore, we can derive the two Holevo quantities $I(\mathbf{E},X)=S(%
\mathbf{E})-S(\mathbf{E}|X)$ and $I(\mathbf{E},Y)=S(\mathbf{E})-S(\mathbf{E}%
|Y)$. By subtracting these from Alice and Bob's mutual information $I(X,Y)$,
we finally get the two key rates in direct and reverse reconciliation, i.e.,
$K(Y|X)$ and $K(X|Y)$.

\begin{figure}[ptbh]
\vspace{-0.1cm}
\par
\begin{center}
\includegraphics[width=0.45\textwidth] {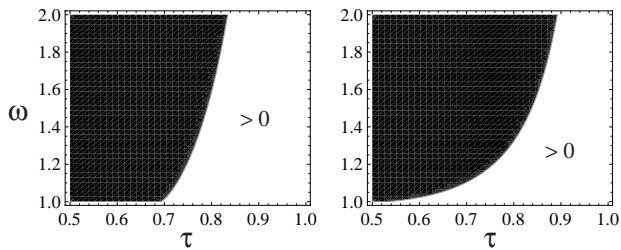}
\end{center}
\par
\vspace{-0.6cm}
\caption{\textit{Left panel}. Rate $K(Y|X)$ in direct reconciliation, as a
function of channel transmissivity $\protect\tau$ and thermal variance $%
\protect\omega$. $K$ is positive in the white area, while it is zero in the
black area. \textit{Right panel}. Rate $K(X|Y)$ in reverse reconciliation,
as function of $\protect\tau$ and $\protect\omega$. White area ($K>0$) is
wider at low $\protect\omega$.}
\label{soglie}
\end{figure}

It is easy to check the existence of wide range of parameters for which
these two rates are strictly positive, so that Alice and Bob can extract a
secret key despite the absence of entanglement (at the input state $\rho_{Aa}
$ and, therefore, also at the output state $\rho_{AB}$). As an example, we
may consider the maximum correlation value $g=\mu-1$ for the separable
Gaussian state $\rho_{Aa}$, and we may take the large modulation limit $\mu
\rightarrow+\infty$, as typical in continuous variable QKD. In this case, we
get the following asymptotical expression for Alice and Bob's mutual
information%
\begin{equation*}
I(X,Y)\rightarrow\log_{2}\frac{\tau\mu}{1+3\tau+(1-\tau)\omega}+O(\mu^{-1}),
\end{equation*}
and the following asymptotical spectra%
\begin{align*}
\nu_{\mathbf{E}}^{-} & \rightarrow(1-\tau)\mu+\tau\omega+O(\mu^{-1}), \\
\nu_{\mathbf{E}}^{+} & \rightarrow\omega+O(\mu^{-1}), \\
\nu_{\mathbf{E}|Y}^{+} & \rightarrow\frac{1+(1-\tau)\omega}{\tau}+O(\mu
^{-1}), \\
\nu_{\mathbf{E}|X}^{\pm} & \rightarrow\xi_{\pm}+O(\mu^{-1}),
\end{align*}
where
\begin{align*}
\xi_{\pm} & :=\frac{\sqrt{(\omega+3)^{2}+\tau^{2}(\omega-3)^{2}-2\tau
(\omega^{2}+7)}}{2} \\
& \pm\frac{(1-\tau)(\omega-3)}{2}.
\end{align*}
Then, using the expansion $h(x)\simeq\log_{2}(ex/2)+O(1/x)$ for large $x$,
we can write the two asymptotical rates%
\begin{align*}
K(Y|X) & =R(\tau,\omega)+h(\xi_{+})+h(\xi_{-}), \\
K(X|Y) & =R(\tau,\omega)+h\left[ \frac{1+(1-\tau)\omega}{\tau}\right] ,
\end{align*}
where we have introduced the common term%
\begin{equation*}
R(\tau,\omega):=\log_{2}\frac{2\tau}{e(1-\tau)[1+3\tau+(1-\tau)\omega ]}%
-h(\omega).
\end{equation*}
As we can see from Fig.~\ref{soglie}, there are wide regions of positivity
for these rates.

In particular, for a pure loss channel ($\omega=1$), the previous
asymptotical rates simplify to the following%
\begin{equation*}
K(Y|X)=\log_{2}\frac{\tau}{e(1-\tau^{2})}+h(3-2\tau),
\end{equation*}
which is positive for any $\tau>0.693$, and%
\begin{equation*}
K(X|Y)=\log_{2}\frac{\tau}{e(1-\tau^{2})}+h\left( \frac{2}{\tau}-1\right) ,
\end{equation*}
which is positive for any $\tau>0.532$.

\section{Discord bound can be tight}

Here we discuss a typical scenario where the optimal backward rate $%
K(\blacktriangleleft)$ of an ideal QKD protocol is exactly equal to the
output discord $D(B|A)$ shared by Alice and Bob. This happens in continuous
variable QKD, where reverse reconciliation is important for its ability to
beat the 3dB loss-limit affecting direct reconciliation~\cite{RMP2}.

Consider an ideal QKD protocol which is based on the distribution of an EPR\
state $\rho_{Aa}$, with CM $\mathbf{V}_{Aa}=\mathbf{V}(\mu)$ defined
according to Eq.~(\ref{Vomega}) with $\mu\geq1$. By performing a rank-1
Gaussian POVM on mode $A$, Alice remotely prepares an ensemble of
Gaussianly-modulated pure Gaussian states on the other mode $a$. For
instance, heterodyne prepares coherent states, while homodyne prepares
squeezed states. On average, mode $a$ is described by a thermal state with
CM $\mu\mathbf{I}$.

Suppose that signal mode $a$ is subject to a pure-loss channel. This means
that Eve is using a beam splitter of transmissivity $\tau$ mixing the signal
mode with a vacuum mode $e$. At the output of the beam splitter, mode $B$ is
detected by Bob, while mode $E$ is stored in a quantum memory coherently
detected by Eve (this is a collective entangling cloner attack with $%
\omega=1 $).

Since the average state of mode $a$\ is thermal and mode $e$ is in the
vacuum, no entanglement can be present between the two output ports $B$ and $%
E$ of the beam splitter. This implies that their entanglement of formation
must be zero $E_{f}(B,E)=0$ and, therefore, the optimal backward rate $%
K(\blacktriangleleft )$ must be equal to the discord $D(B|A)$ of the
Gaussian state $\rho_{AB}$. Since this output state has CM%
\begin{equation*}
\mathbf{V}_{AB}=\left(
\begin{array}{cc}
\mu\mathbf{I} & \sqrt{\tau(\mu^{2}-1)}\mathbf{Z} \\
\sqrt{\tau(\mu^{2}-1)}\mathbf{Z} & (\tau\mu+1-\tau)\mathbf{I}%
\end{array}
\right) ,
\end{equation*}
its discord is easy to compute and is equal to~\cite{Bconj}
\begin{equation*}
D(B|A)=h(\mu)-h[\tau+(1-\tau)\mu].
\end{equation*}
For large modulation ($\mu\rightarrow+\infty$), we have the asymptotic
expression
\begin{equation*}
K(\blacktriangleleft)=D(B|A)=\log_{2}\left( \frac{1}{1-\tau}\right) ,
\end{equation*}
which is positive for any $0<\tau<1$. One can check that this rate can be
achieved by heterodyne detections at Bob's side (and coherent detection at
Alice's side).


\begin{thebibliography}{99}
\bibitem{Nielsen} M.M. Wilde,\textit{\ Quantum Information Theory}
(Cambridge University Press, Cambridge, 2013).

\bibitem{Ekert} A.K. Ekert, Phys. Rev. Lett. \textbf{67}, 661 (1991).

\bibitem{Tele} C.H. Bennett, G. Brassard, C. Crepeau, R. Jozsa, A. Peres,
W.K. Wootters, Phys. Rev. Lett. \textbf{70}, 1895 (1993).

\bibitem{Shor} P. Shor, SIAM J. Comput. \textbf{26}, 1484 (1997).

\bibitem{Qdiscord} H. Ollivier and W.H. Zurek, Phys. Rev. Lett. \textbf{88},
017901 (2001).

\bibitem{Qdiscord2} L. Henderson and V. Vedral, J. Phys. A \textbf{34}, 6899
(2001).

\bibitem{VedralRMP} K. Modi, A. Brodutch, H. Cable, T. Paterek, and V.
Vedral, Rev. Mod. Phys. \textbf{84}, 1655-1707 (2012).

\bibitem{Demon} W.H. Zurek, Phys. Rev. A \textbf{67}, 012320 (2003).

\bibitem{Merging1} D. Cavalcanti, L. Aolita, S. Boixo, K. Modi, M. Piani, A.
Winter, Phys. Rev. A \textbf{83}, 032324 (2011).

\bibitem{Merging2} V. Madhok and A. Datta, Phys. Rev. A \textbf{83}, 032323
(2011).

\bibitem{Remote} B. Dakic, Y.O. Lipp, X. Ma, M. Ringbauer, S. Kropatschek,
S. Barz, T. Paterek, V. Vedral, A. Zeilinger, C. Brukner, and P. Walther,
Nature Phys. \textbf{8}, 666-670 (2012).

\bibitem{Qmetrology} D. Girolami, T. Tufarelli, and G. Adesso, Phys. Rev.
Lett. 110, 240402 (2013).

\bibitem{Gu} M. Gu, H.M. Chrzanowski, S.M. Assad, T. Symul, K. Modi, T.C.
Ralph, V. Vedral, and P. Koy Lam, Nature Phys. \textbf{8}, 671-675 (2012).

\bibitem{Qillumination} C. Weedbrook, S. Pirandola, J. Thompson, V. Vedral,
and M. Gu, \textit{Discord Empowered Quantum Illumination}, preprint
arXiv:1312.3332.

\bibitem{Gisin} N. Gisin, G. Ribordy, W. Tittel, and H. Zbinden, Rev. Mod.
Phys. \textbf{74}, 145 (2002).

\bibitem{Filip} R. Filip, Phys. Rev. A \textbf{77}, 022310 (2008).

\bibitem{Usenko} V.C. Usenko and R. Filip, Phys. Rev. A \textbf{81}, 022318
(2010).

\bibitem{Weedbrook2010} C.~Weedbrook, S.~Pirandola, S.~Lloyd, and
T.C.~Ralph, Phys. Rev. Lett. \textbf{105}, 110501 (2010).

\bibitem{Weedbrook2012} C.~Weedbrook, S.~Pirandola, and T.C.~Ralph, Phys.
Rev. A \textbf{86}, 022318 (2012).

\bibitem{Weedbrook2014} C. Weedbrook, C. Ottaviani, and S. Pirandola, Phys.
Rev. A \textbf{89}, 012309 (2014).

\bibitem{Renner2005} R.~Renner, N.~Gisin, and B.~Kraus, Phys. Rev. A \textbf{%
72}, 012332 (2005).

\bibitem{Pirandola2009} S.~Pirandola, R.~Garc\'{\i}a-Patr\'{o}n,
S.L.~Braunstein, and S.~Lloyd, Phys. Rev. Lett. \textbf{102}, 050503 (2009).

\bibitem{RMP} C. Weedbrook, S. Pirandola, R. Garcia-Patron, N.J. Cerf, T.C.
Ralph, J.H. Shapiro, and S. Lloyd, Rev. Mod. Phys. \textbf{84}, 621 (2012).

\bibitem{SideCH} S.L. Braunstein and S. Pirandola, Phys. Rev. Lett. \textbf{%
108}, 130502 (2012).

\bibitem{Nota2} Variables can be discrete or continuous. In the latter case,
sums become integrals and probability distributions become densities.

\bibitem{Winter} I. Devetak and A. Winter, IEEE Trans. Inform. Theory.
\textbf{50}, 3183 (2004).

\bibitem{Qcap} B. Schumacher and M.A. Nielsen, Phys. Rev. A \textbf{54},
2629 (1996).

\bibitem{Qcap2} S. Lloyd, Phys. Rev. A \textbf{55}, 1613 (1997).

\bibitem{Koashi} M.~Koashi and A.~Winter, Phys. Rev. A \textbf{69}, 022309
(2004).

\bibitem{Renner} R. Renner, Nature Phys. \textbf{3}, 645 (2007).

\bibitem{Renner2} R. Renner, and J.I. Cirac, Phys. Rev. Lett. \textbf{102},
110504 (2009).

\bibitem{Noteiso} Up to isometries which do not increase Eve's information.

\bibitem{Add} Note that we can always add another ancilla $E^{\prime}$ to
make the output $\Psi_{ABEE^{\prime}P}$ globally pure.

\bibitem{cohe} This reasoning can easily be extended to consider coherent
detections for Eve and Alice.

\bibitem{B92} C.H. Bennett, Phys. Rev. Lett. \textbf{68}, 3121-3124 (1992).

\bibitem{Notation1} Note that, for $A$,$B$ quantum systems and $X,Y$
classical variables, we use $I(X,Y)$ for the classical mutual information, $%
I(A,B)$ for the quantum mutual information, and $I(A,X)$ for the Holevo
information.

\bibitem{DW1} I. Devetak and A. Winter, Phys. Rev. Lett. \textbf{93}, 080501
(2004).

\bibitem{DW2} I. Devetak and A. Winter, Proc. R. Soc. Lond. A \textbf{461},
207-235 (2005).

\bibitem{DWnote} In general, the DW rate satisfies Eq.~(\ref{DWdr}) with $%
\geq$, the separation ($>$) being possible for higher-rank POVMs~\cite%
{DW1,DW2}. These POVMs are considered here by including their detection
noise in system $P$ (e.g., as in Fig.~\ref{PicD}). What remain are rank-1
POVMs applied to $A $ and $B$.

\bibitem{RevCOH} R. Garc\'{\i}a-Patr\'{o}n, S. Pirandola, S. Lloyd, J.H.
Shapiro, Phys. Rev. Lett. \textbf{102}, 210501 (2009).


\end{thebibliography}

\begin{thebibliography}{}
\bibitem{RMP2} C. Weedbrook, S. Pirandola, R. Garcia-Patron, N.J. Cerf, T.C.
Ralph, J.H. Shapiro, and S. Lloyd, Rev. Mod. Phys. \textbf{84}, 621 (2012).

\bibitem{twomodes} S. Pirandola, A. Serafini, and S. Lloyd, Phys. Rev. A
\textbf{79}, 052327 (2009).

\bibitem{VeriDISCO} S. Rahimi-Keshari, C.M. Caves, and T.C. Ralph, Phys.
Rev. A \textbf{87}, 012119 (2013).

\bibitem{colleGAUSS} S. Pirandola, S.L. Braunstein, and S. Lloyd, Phys. Rev.
Lett. \textbf{101}, 200504 (2008).

\bibitem{Bconj} S. Pirandola \textit{et al}., preprint arXiv:1309.2215.
\end{thebibliography}
\end{document}